# REFLECTION ON TRAINING, EXPERIENCE, AND INTRODUCTORY STATISTICS A MINI-SURVEY OF TERTIARY LEVEL STATISTICS INSTRUCTORS

Rossi A. Hassad
Mercy College, United States
Rhassad@mercy.edu

*Instructors of statistics who teach non-statistics majors possess varied academic backgrounds, and hence it is reasonable to expect variability in their content knowledge, and pedagogical approach. The aim of this study was to determine the specific course(s) that contributed mostly to instructors' understanding of statistics. Courses reported were described as advanced or graduate level, and classified as application-based, math, multivariate, probability, and research. The majority, 9 (56%) attributed their understanding of statistics to either an application-based or research course, and of those, 7 (44%) reported negative feelings about their introductory courses. These findings underscore the importance of authentic activities, and constructivist pedagogy toward facilitating statistical literacy. Research is needed to determine the effect of instructors' academic preparation on their knowledge, attitudes, and practices.*

## INTRODUCTION

Instructors who teach statistics (especially introductory courses) to non-statistics majors possess diverse academic backgrounds, including psychology, behavioral sciences, education, sociology, mathematics, engineering science, biostatistics, statistics (mathematical and applied), epidemiology, economics, and public health. This heterogeneity in training and preparation is likely to equip them with varying degrees of knowledge and skills in statistics (content knowledge), which raises the following questions. How does this knowledge base translate into pedagogical content knowledge (Shulman, 1987), teaching, and student performance? Also, which specific course or other exposure from training or practice most of all facilitated their understanding of statistics, and how does this relate to their cognitive style (Martinsen and Kaufmann, 1999; Lovett and Greenhouse, 2000) and pedagogical approach?

These questions are relevant to effective teaching and learning, as instructors may be inclined to teach the way they learned (Rusley, 2003). And this can be counterproductive when there is a mismatch of learning styles (Kolb, 1984; Fielding, 1994), which is not recognized and addressed by the instructor. This is particularly relevant to statistical methods, which have underlying concepts and assumptions that may be viewed as difficult and counterintuitive (such as aspects of probability theory and hypothesis testing).

## THE TRAINING AND RESEARCH GAP

The statistics education literature is replete with best practices for facilitating quantitative reasoning (Lovett and Greenhouse, 2000), and these strategies have become the focus of professional development programs for statistics instructors. A major project in this regard was STATS: Statistical Thinking with Active Teaching Strategies (Rossman, 1996-1999) which conducted workshops for mathematicians who teach statistics but have little formal training in the subject. The target audience for these workshops has since been expanded to include instructors of the introductory statistics course (Pearl and Short, 2005).

Not adequately addressed in these training programs is the diversity of learning styles and learning strategies among students, which is considered the primary challenge to implementing effective pedagogy. For example, algorithmic learners are inclined to initially show resistance to reform-based teaching which emphasizes conceptual thinking and understanding rather than mathematical underpinnings and procedural knowledge. This barrier must be resolved before meaningful and deep learning can occur. Otherwise we may lose many potential educators and practitioners. They will fail or drop the course, and go on to either change their academic major or not complete their degree. Also, there are those students who may have had an unpleasant experience with the course, struggled and managed to pass, but are discouraged from taking another statistics course, or engaging in the discipline.

SELECTED STRATEGIES (FOR BRIDGING THE GAP)

Toward promoting meaningful learning, concept mapping (Novak, 1991; Verkoeijen *et al.*, 2002) is considered effective and efficient. It is favorable to a broad spectrum of cognitive styles, and enables students to identify, integrate, and apply course concepts (Jonassen, 1996). Concept maps are primarily semantic networks expressing the interrelationship among concepts within a domain of information. The exercise of creating this map helps students to make connections between theory and practice, and build on existing knowledge. Furthermore, concept maps aid the instructor in assessing understanding. Indeed, the key implication here is that the instructor should be conversant with learning theories and their applications, however, this is more the exception rather than the norm.

Besides focusing on teaching methodology, introductory statistics courses with central and unifying themes such as "variability" (Moore, 1997), "prediction" (McLean, 2000), "data management," and "decision-making" (Hassad, 2002) have been suggested as effective for facilitating statistical thinking and literacy (Wild and Pfannkuch, 1999; Chance, 2002). Toward this end, the epidemiological model has been posited as a practical framework for designing introductory statistics courses to achieve quantitative reasoning (Stroup *et al.*, 2004). This is especially applicable to the evidence-based disciplines (such as health and behavioral sciences).

OBJECTIVE

The primary objective of this mini-survey was to ascertain from instructors and practitioners of statistics, the specific course(s) or other exposure which contributed mostly to their understanding and application of statistics.

METHODOLOGY

In January 2005, an exploratory mini email survey was conducted via ALLSTAT and SMRSNET. Two broad open-ended (and complementary) questions were asked.
(1) Which course(s) or other exposure contributed mostly to your understanding and application of statistics?
(2) Do you recall when you first said any of the following? ("*aha" experiences*)
(a) Oh, I see, this is how it works! (b) It's all coming together and making sense! (c) I got it! (d) This is so cool! (e) I like this!

RESULTS

Sixteen (16) responses were received. Respondents were either college instructors of statistics or statisticians/data analysts with teaching experience. All courses were described as *advanced* or *graduate* level only, and based on thematic analysis, were classified as follows:

| Type of course which instructors reported as being most instrumental to their understanding of statistics (*n*=16) | |
|---|---|
| Course Description | Number |
| Application-based* | 7 |
| Research | 2 |
| Multivariate | 2 |
| Probability | 2 |
| Math | 3 |

*Actuarial Science, Biometry, Business, Ecology, Economics, Psychology, Thesis, Dissertation, and SPSS. A few respondents listed multiple courses. Courses reflect "aha" experiences.*

The seven (7) respondents who reported "application-based" courses, expressed (in their responses) negative feelings (see below) about their introductory and earlier statistics courses.
"My introduction to statistics was an incomprehensible course…."
It "wash[ed] over me like a wave".
"I didn't even catch the nuance between the two [standard error and standard deviation]. I just used them interchangeably."

"The limit was the actual calculations since no computers were available…."
"I didn't think the course was that interesting…"
"Grueling"
I "didn't really learn data analysis…."
"Statistics was just another math course. I never saw the problem solving promise of the field."

DISCUSSION

The majority of the instructors/practitioners, nine (56%) attributed their understanding of statistics to either an advanced or graduate level application-based or research course (see table). Of these, seven (44%) reported or implied that their introductory statistics course was not helpful, as it was too math-oriented, and not practical. This profile seems to characterize instructors who are predisposed to active teaching and learning strategies (such as cooperative and problem-based learning). On the other hand, respondents who reported math, probability and multivariate courses, may possess varied cognitive styles and pedagogical approaches, as these courses can be viewed as stimuli for different types of learning, understanding and reasoning.

In particular, multivariate statistics is more consistent (than mathematics and probability) with real-world phenomena. Accordingly, it can be argued that multivariate statistics is akin to application-based courses, and is more likely to lead to conceptual understanding and deep learning. Multivariate statistical methods can facilitate understanding of concepts such as the biopsychosocial framework of disease, and multiple causality, which are the predominant models in the health and behavioral sciences. It would be worthwhile to find out what type of understanding and reasoning was engendered by each course classification (see table), and the resulting pedagogical approach. In this regard, the learning context as well as cumulative learning (rather than a discrete course) must be considered.

The delayed understanding of statistics is disconcerting. For all, this came with advanced (*not introductory*) courses only. This concern has been a focus of the introductory statistics education reform movement for more than a decade (Garfield *et al.*, 2002), however, despite the achievements, much remains to be done. A more systematic and evidence-based approach is required. This finding underscores the importance of authentic learning activities (Rossman, 1997; Engel, 2002; Reeves *et al.*, 2002), within the broader constructivist pedagogy (Garfield, 1993; Verkoeijen *et al*., 2002; Seipel and Apigian, 2005) in promoting statistical thinking and reasoning.

Improvement of the introductory statistics course is necessary so that students can emerge with useful and transferable knowledge and skills. Instructors should use material that is *R*elevant, *I*nteresting, and *S*imple, and *K*ind (RISK) so that students can meaningfully experience the concepts rather than be victims of passive learning. Instructors (especially in the health and behavioral sciences) should adopt the epidemiological model to guide course development, as this encompasses the themes of variability, prediction, data management, and decision-making, with reference to real-world data, as well as salient and universal issues (health). This can motivate students to explore data, and discover meaning. Moreover, statistics is fast becoming a graduation requirement for most majors, and especially in psychology, it is regarded by some academics as "the single most important course in terms of admittance into graduate schools" (Alder and Vollick, 2000).

Notwithstanding the methodological limitations of this mini-survey (sample size, design, recall bias, external validity, etc.) these observations are plausible, and should be further explored with a larger and more scientific study (with attention to personal, socio-demographic, and contextual factors). Also, large-scale research is needed to determine the effect of instructors' academic training and professional preparation on their knowledge, attitudes, conceptions, and pedagogical practices in the context of teaching statistics, in particular, the introductory and basic courses for non-statistics majors.